\documentstyle[aps,epsf,preprint]{revtex}
\newcommand{\pspicture}[1]{
\centerline{\setlength\epsfxsize{9.2cm}\epsfbox{#1}}}
\newcommand{\pstwopictures}[2]{
\centerline{
\hspace{-20pt}
\epsfxsize=225pt
\epsfbox{#1}
\hspace{-20pt}
\epsfxsize=225pt
\epsfbox{#2}
}}
\ifx\DeclareFontShape\undefined
\else
        \DeclareSymbolFont{lasy}{U}{lasy}{m}{n}
        \SetSymbolFont{lasy}{bold}{U}{lasy}{b}{n}
        \let\Box\undefined
        \DeclareMathSymbol\Box{0}{lasy}{"32}
\fi

\ifx\epsffile\undefined
\else
\fi
\begin{document}

\title{
\begin{flushright}
{\small
GUTPA 95--10--4 \\
EDINBURGH 96/8\\
LTH 380 \\
SHEP 96/19  \\
hep-lat/9608063
}
\end{flushright}
{
\Huge
The effect of tree-level and mean-field improvement
on the light-hadron spectrum in quenched QCD.}
}

\author{UKQCD Collaboration\footnote{H.Shanahan@physics.gla.ac.uk}}

\author{H.P.~Shanahan, C.T.H.~Davies}
\address{Department of Physics \& Astronomy, University of Glasgow,
Glasgow, G12 8QQ, Scotland, U.K.}

\author{K.C.~Bowler, R.D.~Kenway, 
D.G.~Richards\footnote{Address until September 30,
HEP Division, Bldg. 362, Argonne National Laboratory, Argonne, IL
60439, USA},
P.A.~Rowland, S.M.~Ryan}
\address{Department of Physics \& Astronomy, The University of Edinburgh,
Edinburgh EH9~3JZ, Scotland, U.K.}

\author{P.~Lacock, C.~Michael}
\address{Theoretical Physics Division, Department of Mathematical Sciences,
University
of Liverpool, Liverpool L69 3BX, U.K.}

\vskip2truecm

\author{D.R.~Burford, N.~Stella}
\address{Department of Physics, University of Southampton, Southampton
SO17~1BJ, U.K.}

\author{H.~Wittig}
\address{DESY-IfH, Zeuthen, Platanenalle 6, D-15738 Zeuthen, Germany}

\maketitle


\begin{abstract}
We compute the light hadron mass spectrum at $\beta=5.7$ using the
$O(a)$-improved Sheikholeslami-Wohlert (SW) fermion action with two
choices of the clover coefficient: the classical value, $c=1$, and a
mean-field or tadpole-improved estimate $c=1.57$.  We compare our
results with those of the GF11 Collaboration who use the Wilson fermion
action ($c=0$).

We find that changing $c$ from zero to 1 and 1.57 leads to significant
differences in the masses of the  chirally extrapolated and strange
pseudoscalar and vector mesons, the nucleon, the $\Delta$,  and also in
the Edinburgh plot. A number of other  quantities, for example $m_V^2 -
m_{PS}^2$, $J$, $am_K/am_\rho$ and  $am_{K^*}/am_\rho$
 do not appear to
change significantly.

We also investigate the effect of changing the lattice volume from
approximately $(2 \, \mbox{fm})^3$ to $(2.6 \, \mbox{fm})^3$. We find
that the meson masses are consistent to within  one standard deviation
and baryon masses are consistent to within two standard deviations.

\end{abstract}


\section{Introduction}
The {\it ab initio} calculation of the light hadron spectrum is a major goal of
lattice QCD.  A calculation of the light-hadron
spectrum giving results in good agreement with experiment would be a
demonstration that QCD describes long-distance strong-interaction
physics.  Furthermore, the calculation is an essential precursor to the
calculation
of other non-perturbative observables in QCD, such as $B_K$, $B_B$,
leptonic and semi-leptonic decay matrix elements and the moments of
the nucleon structure function.  Lattice calculations are
however subject to
systematic errors from the non--zero lattice spacing, the finite volume of
the lattice, the extrapolation in the valence quark mass to the chiral
limit, and the quenched approximation.  In this paper, the effects of
the first two sources of error will be examined.

Symanzik \cite{symanzik} proposed an improvement programme for reducing
the dependence of observables on the lattice spacing, $a$, by adding
to the action higher-dimension operators with
appropriately
calculated coefficients.  This should enable a more reliable
extrapolation to the continuum limit, using data at larger values of
the lattice spacing.  Given that the computational effort scales as
$a^{-6}$ in the quenched approximation, the potential savings are
considerable.

The standard gluon action has discretisation errors of $O(a^2)$.
The Wilson fermion action, on the other hand, has discretisation
errors of $O(a)$.  Therefore, the first step in the Symanzik
improvement programme is to reduce the leading-order error of the
fermion action to the same order as that of the gluon action.  The
resulting Sheikholeslami-Wohlert (SW) action~\cite{sheikwo} introduces
an extra operator, $P(x)$, the so-called clover term, to the
original action, multiplied by a parameter $c$:
\begin{equation}
S^F_{\rm SW} = S^F_{\rm W}(\kappa, r) + a^4  \, c \kappa r  \, \sum_x
\overline{\psi}_x P(x) \psi_x \; ,
\label{eq:clover-action}
\end{equation}
where $S^F_W(\kappa, r)$ is the standard Wilson action defined as
\begin{eqnarray}
S^F_{\rm W}(\kappa, r) = a^4 \sum_x\left\{
\overline{\psi}_x \psi_x  \right. &+&
\left. \kappa \sum_\mu \left( \overline{\psi}_x\left( \gamma_\mu - r \right) U_\mu(x)
\psi_{x + \hat{\mu}} \right. \right. \nonumber \\
& & \left.  \;\;\;\;\;\;\;\;\; -  \;
 \left. \overline{\psi}_{x + \hat{\mu}} \left( \gamma_\mu + r \right) U^\dagger(x) \psi_x
\right) \right\}  \; ,
\end{eqnarray}
and
\begin{equation}
P(x) = \frac{-ia}{2} \sum_{\mu,\nu} F^c_{\mu\nu}(x) \sigma_{\mu\nu} \; ,
\end{equation}
$F^c_{\mu\nu}(x)$ is a lattice definition of the field strength
tensor, detailed in \cite{ukqcd:npblight}.

There is a value of the parameter $c$, $c_{\rm non-pert}$, which
removes all $O(a)$ errors from spectral
quantities~\cite{luwe,alpha}.  In this paper, we compare the
spectrum obtained using the Wilson fermion action ($c=0$) with that
obtained using the SW fermion action with two choices of $c$: the
classical value, $c=1$, and a mean-field or tadpole-improved estimate of
$c_{\rm non-pert}$. Other approaches to improvement are described in
refs.~\cite{edwards,alford,morningstar,woloshyn,niedermeyer}.


%
The tadpole--improved estimate of $c$ is obtained
following Lepage and Mackenzie~\cite{kermit}
by replacing the gauge links, $U_\mu(x)$ by
\begin{equation}
\tilde{U}_\mu(x) =\frac{1}{u_0}U_\mu(x) \; .
\label{eq:tadpole-gauge}
\end{equation}
We choose
\begin{equation}
u_0 = \langle \frac{1}{3}{\rm Tr} U_\Box \rangle^{\frac{1}{4}}.
\end{equation}
Consequently, the effect of tadpole improvement on the SW action is to set
\begin{eqnarray}
c \;\;& =& \;\; \frac{\tilde{c}}{u_0^3}\\
\kappa \;\;& = & \;\; \frac{\tilde{\kappa}}{u_0}.
\label{eq:kreplace}
\end{eqnarray}
Tree--level theory should then provide more reliable estimates of
$\tilde{c}$ and  the critical value of $\tilde{\kappa}$ which we denote
$\tilde{\kappa}_{\mathrm{crit}}$; we take  $\tilde{c}=1$ and expect
$\tilde{\kappa}_{\mathrm{crit}}$ to be close to $\frac{1}{8}$.
This prescription maintains the $O(a)$ improvement
and it is believed that the size
of the remaining discretisation error will be reduced.

The paper is organised as follows.  In the next section we outline the
computational methods. In section III, we
explore three values of the clover coefficient
at $\beta=5.7$ by including the results
from the GF11 collaboration~\cite{GF11}.
The observables studied are:
the $\rho$ and $\pi$ masses, vector pseudoscalar
mass splittings, the $J$ parameter (proposed by Lacock and Michael
\cite{UKQCD:J}),  valence $\overline{s} s$ meson masses,  the spin
$1/2$ and $3/2$ baryon masses and the Edinburgh plot.
A study is also made of possible finite size
effects by computing the spectrum at a smaller lattice volume, using
one value of the clover coefficient.
Finally, in Section IV, we present our conclusions.

\section{Computational Details}

\subsection{Simulation Parameters}

Two lattice sizes, $12^3 \times 24$
and $16^3 \times 32$, at $\beta=5.7$,
were used, with 482 configurations generated on
the former and 142 configurations on the latter.
We used a combination of the over--relaxation (OR)
algorithm \cite{Creutz-gauge} and the Cabbibo--Marinari (CM) algorithm
\cite{CM-gauge}.
The gauge configurations
were separated by 100 compound sweeps, where a
compound sweep is defined as five OR sweeps followed by one CM sweep.
A detailed description of the algorithms used can be found in
\cite{ukqcd:npblight}.

Quark propagators were calculated at two $\kappa$ values.
  These values were chosen so that
the corresponding quark masses  straddle the strange quark mass.
On the larger lattice, propagators were
calculated using both $c=1$ and the tadpole--improved value of
$c=1.57$.
On the smaller lattice, propagators were calculated using
the tadpole--improved value of $c$ only.

To increase the overlap of the operators with the ground
state, all of the propagators were calculated using
both a local source and a Jacobi--smeared source with  r.m.s.\ radius of
$2.2 a$ \cite{ukqcd:smearing}.
Local sinks were used for all
propagators.
The propagators were calculated
using the minimal residual algorithm, which is described in
detail in~\cite{ukqcd:npblight}.

The correlators used to extract the hadron masses are listed in
Table~\ref{operator-list}; for further details see \cite{ukqcd:strange}. We
computed meson correlators using quarks degenerate and non-degenerate
in mass, giving three possible mass combinations for each meson state.
Furthermore, each quark propagator can be either local
or smeared, giving three possible correlators for each mass
combination. However, we computed baryon correlators only for
degenerate quark masses, using either all smeared or all local quark
propagators. Therefore, for each baryon state we have two mass
combinations each with two types of sources.  In order to maximise the
sample size, the discrete time symmetry of the correlators was utilised
and the data for $t \in [0, T/2]$ averaged with the data at $T - t$,
where $T$ is the temporal size of the lattice.

These calculations were performed on the   Meiko
i860 Computing
Surfaces at the Edinburgh Parallel Computing Centre.

\subsection{Fitting}

We have performed
multi-exponential fits of meson correlators to
\begin{equation}
\sum_{\vec{x}}
\left\langle 0 \left|{M(\vec{x},t) M^\dagger(0)}\right| 0 \right\rangle 
=
\sum_{n=0}^{n_{max}} A_n \cosh{(m_n(\frac{T}{2}-t))} \;\; ,
\label{eq:meson-twopt}
\end{equation}
and baryon correlators to
\begin{equation}
\label{eq:baryon-twopt}
\sum_{\vec{x}}
\left\langle 0 \left|{B(\vec{x},t) \overline{B}(0)}\right| 0 \right\rangle 
=
\sum_{n=0}^{n_{max}}
\left( B_n \exp{(-m_n t)} + C_n \exp{(-m_n^P (T-t))} \right) \;\; .
\end{equation}
$B_n$ is the amplitude of the state labelled by $n$, and $C_n$ is that
of the (heavier) parity partner and $n_{max} \ge 1$.

The following criteria for multi--exponential fits
have been used :
\begin{itemize}

\item acceptable values for the quality of fit, Q, and $\chi^2/\rm{d.o.f.}$;
\item stability of the result for the ground state mass;
\item agreement between the result obtained using a single-exponential
fit and a double-exponential fit;
\item ability of of the fitting algorithm to resolve two masses.
\end{itemize}

The variable $Q$, which is a function of $\chi^2$ and $\nu=\rm{d.o.f.}$
 is defined~\cite{num-recipies} as 
\begin{equation}
Q(\nu, \chi^2) \equiv \frac{1}{\Gamma(\nu/2)} \int_{\chi^2/2}^\infty e^{-t} t^{\nu/2 - 1} dt \;\; .
\end{equation}
It represents the probability that given $\nu$ normal, random, uncorrelated
variables, with a mean of $0$ and unit variance, have a sum of squares 
which is greater than $\chi^2$.
An acceptable value for $Q$  lies around $0.5$; a much smaller value
indicates that the model used is incorrect, whereas a value approaching 1
indicates that too many parameters are being used.
A criterion of stability which we used 
is that the mass obtained does not change noticeably when the 
minimum time slice of the fit was changed slightly.
The parameters were determined by minimising the $\chi^2$ using the
Levenberg--Marquardt algorithm \cite{num-recipies,marquardt}.  
Correlations between
all time slices, and types of operator for simultaneous fits, were
included.  The covariance matrix was inverted using Singular Value
Decomposition, without eliminating any eigenvalues.  The bootstrap
algorithm \cite{bootstrap}, using 1000 bootstrap subsamples, was used
to determine the 68\% confidence levels, regenerating the covariance
matrix for each subsample.

\begin{figure}
\pstwopictures{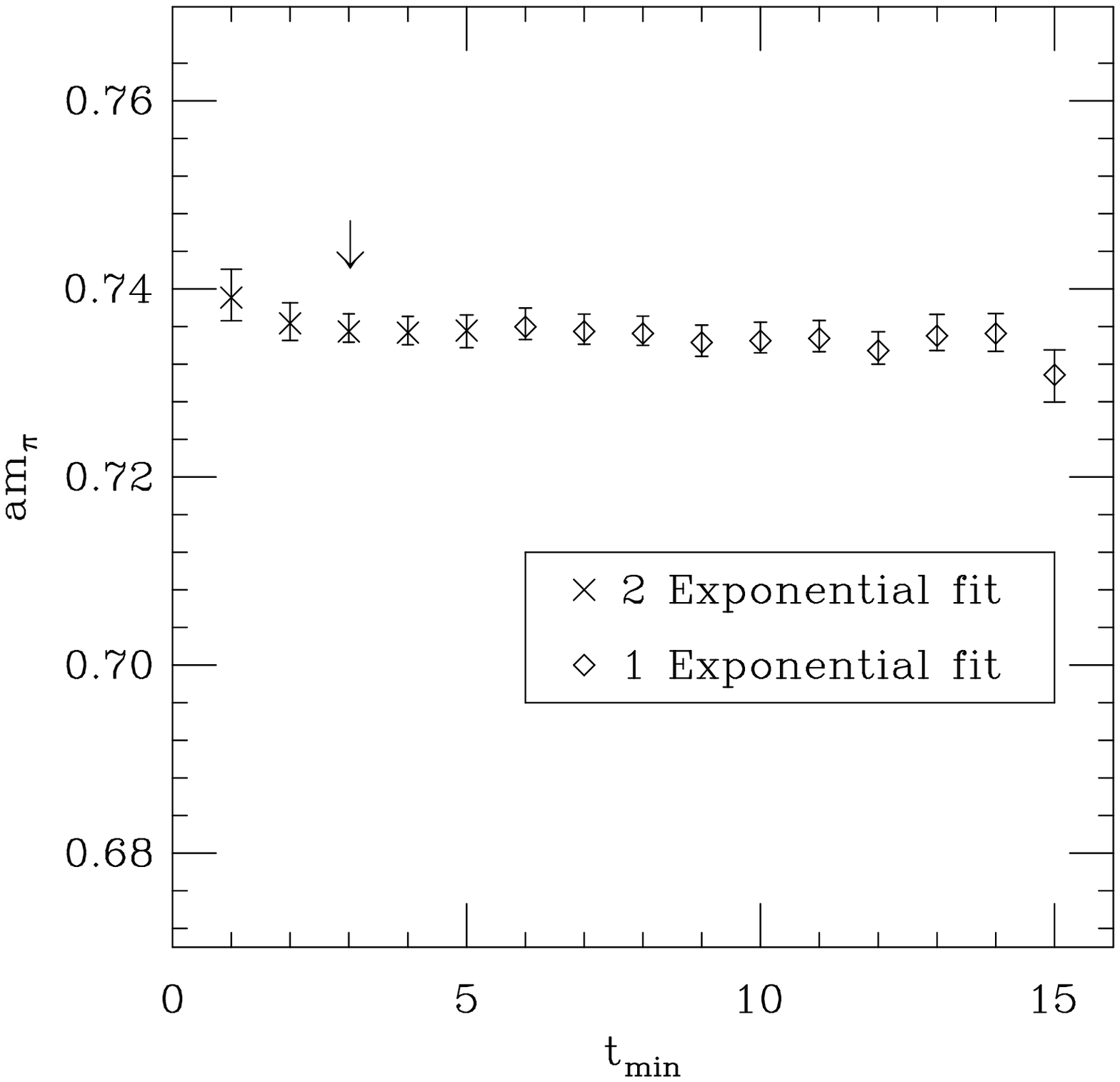}{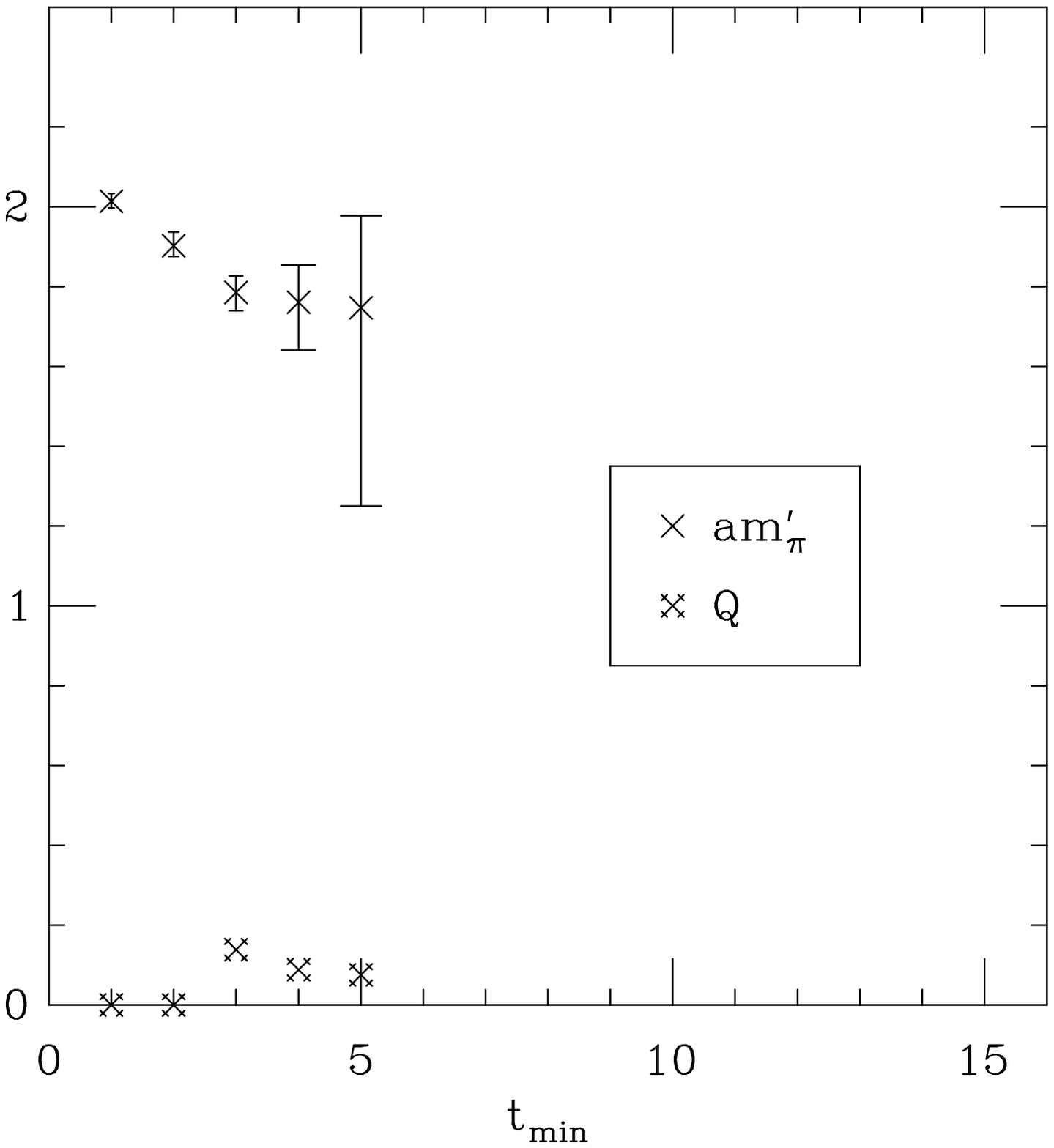}
\caption{$am_{PS}$, $am'_{PS}$ and $Q$  versus
$t_{min}$ for $c=1.57$,  $16^3 \times 32$, $\kappa_1, \kappa_2 =
0.13843$,
using local and smeared propagators.
The arrow indicates which fit range was used for the final result. }
\label{fig:ps-57-multi-fit}
\end{figure}

\begin{figure}
\pstwopictures{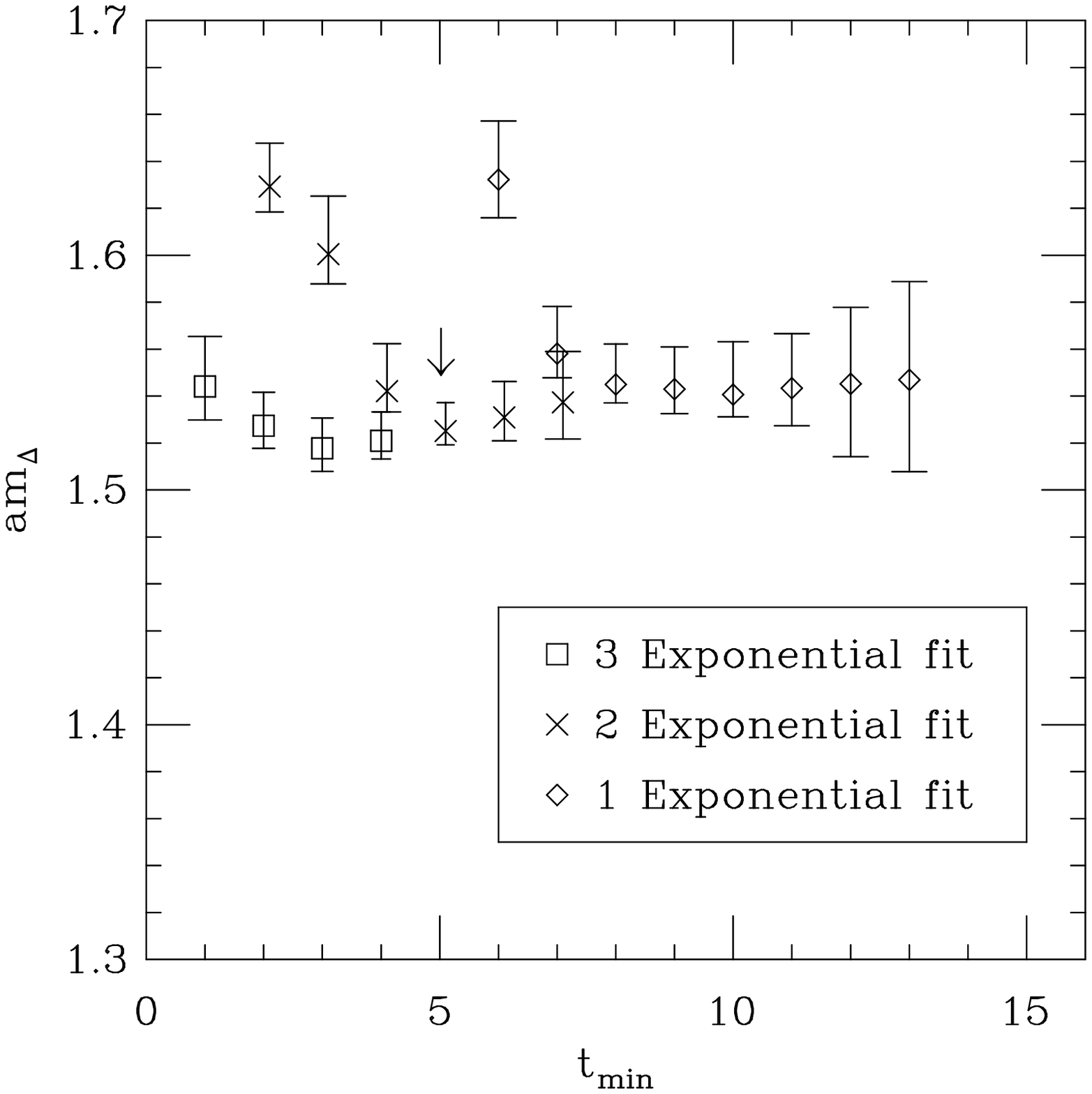}{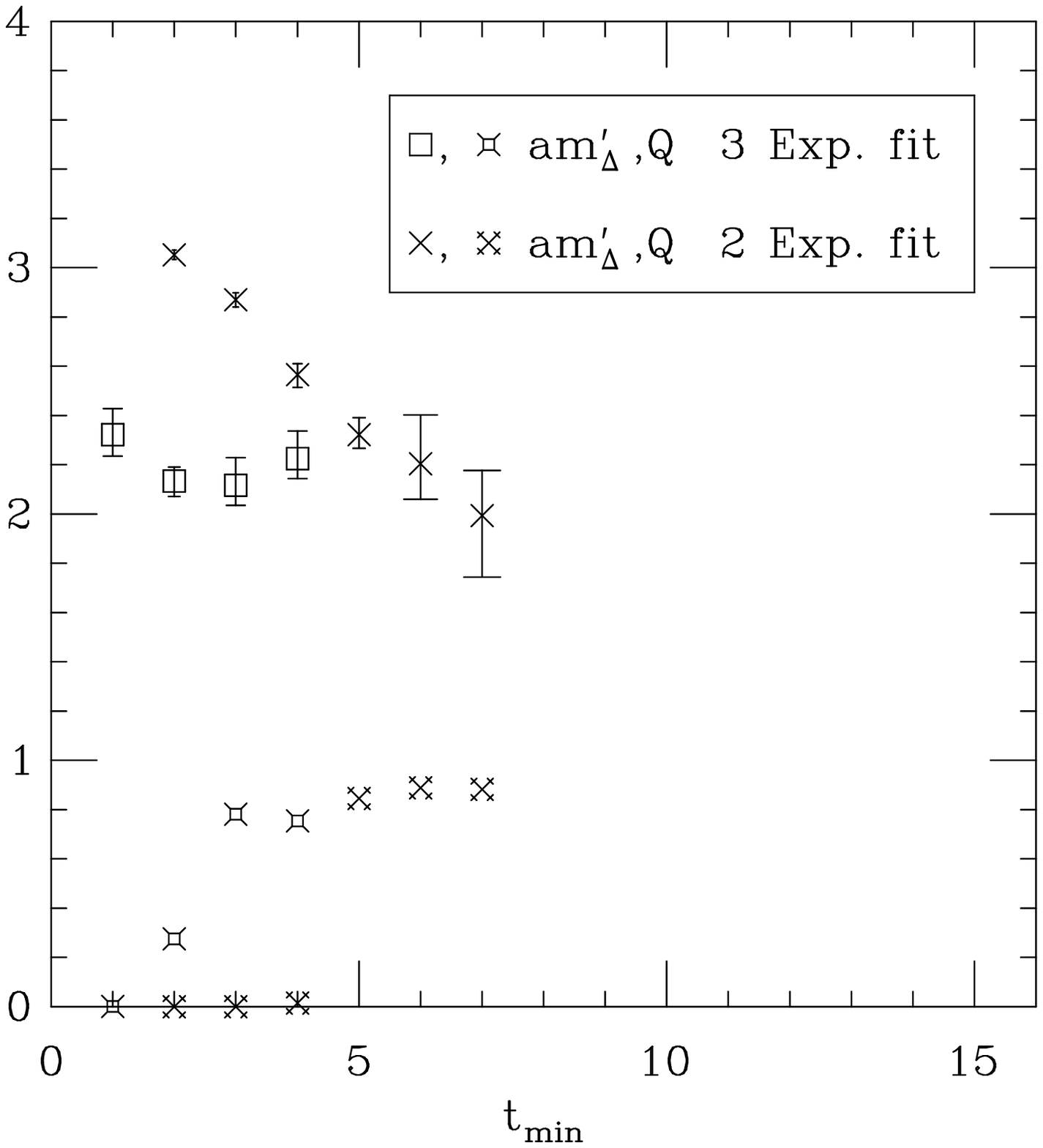}
\caption{$am_{\Delta}$, $am'_{\Delta}$ and $Q$  versus
$t_{min}$ for $c=1.57$, $16^3 \times 32$, $\kappa = 0.13843$,
using  local and smeared propagators.
The arrow indicates which fit range was used for the final result.}
\label{fig:delta-57-multi-fit}
\end{figure}

Examples of the  multi--exponential fits for the pseudoscalar,
vector, nucleon and $\Delta$ are shown in 
Fig.~(\ref{fig:ps-57-multi-fit})  to
Fig.~(\ref{fig:delta-57-multi-fit}). We emphasise that these are {\it not}
effective mass plots, but plots of the mass obtained for a
given fixed $t_{max}$ and varying $t_{min}$.
In obtaining results for
 the smaller lattice, despite having significantly larger
statistics, it was more difficult to satisfy  the
above fit criteria
  than for  the larger lattice.
The pseudoscalar mass was determined using all available smearing
types and a 2--exponential fit. Fit ranges of
3--12 and 3--16 were chosen for the smaller and larger volumes
repectively.
In the case of the vector, the high statistics at the smaller volume
allowed the use of both   $\Gamma$ matrices, listed in
Table~\ref{operator-list}, while for the larger lattice, only
$\vec{V}_1 \;=\; \overline{\psi} \vec{\gamma} \psi $
was used.
All three different smearing types were used in both fits.
Fit ranges of 4--12 and 4--16 were used and a 2--exponential fit.

\begin{figure}
\pstwopictures{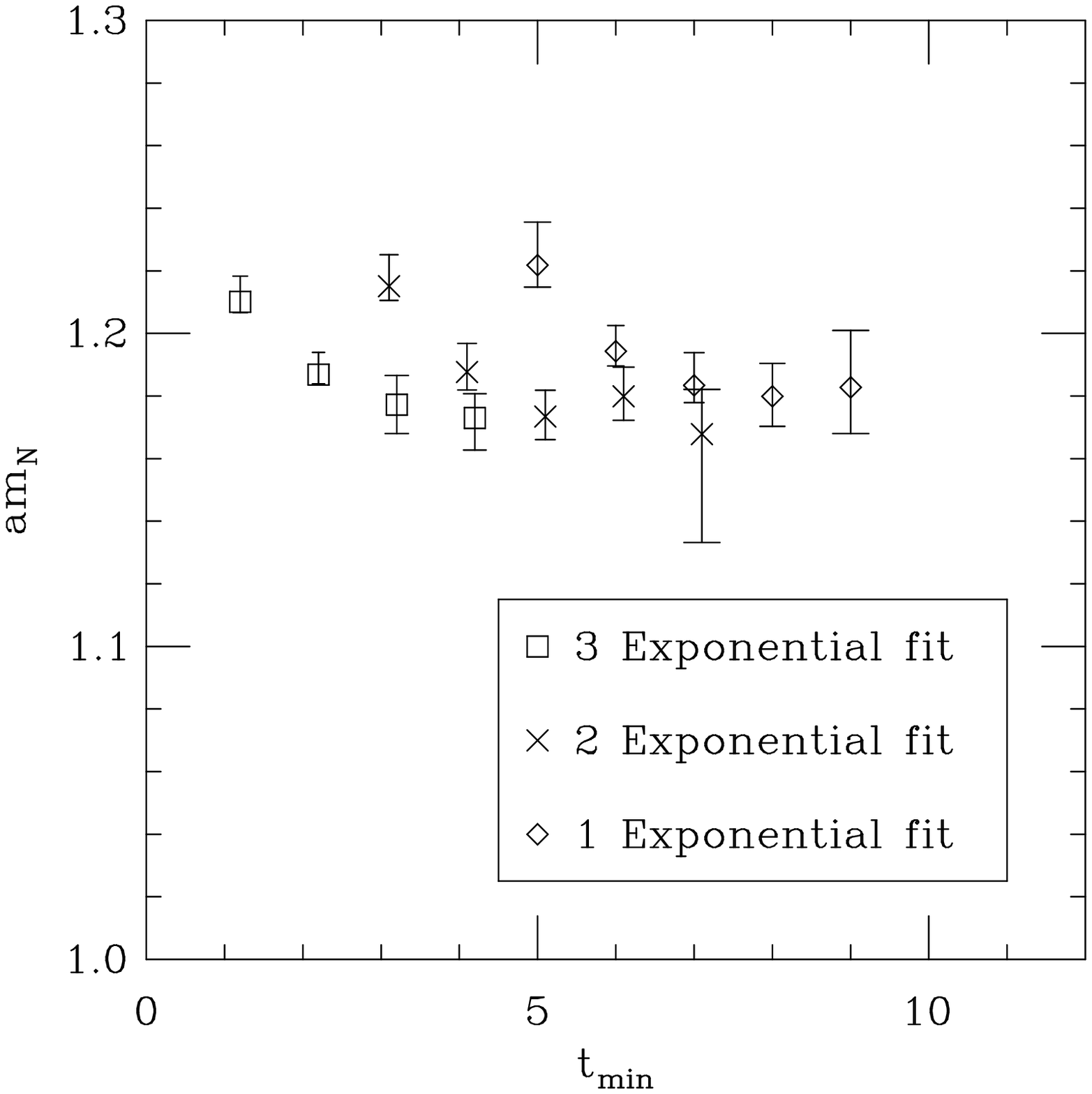}{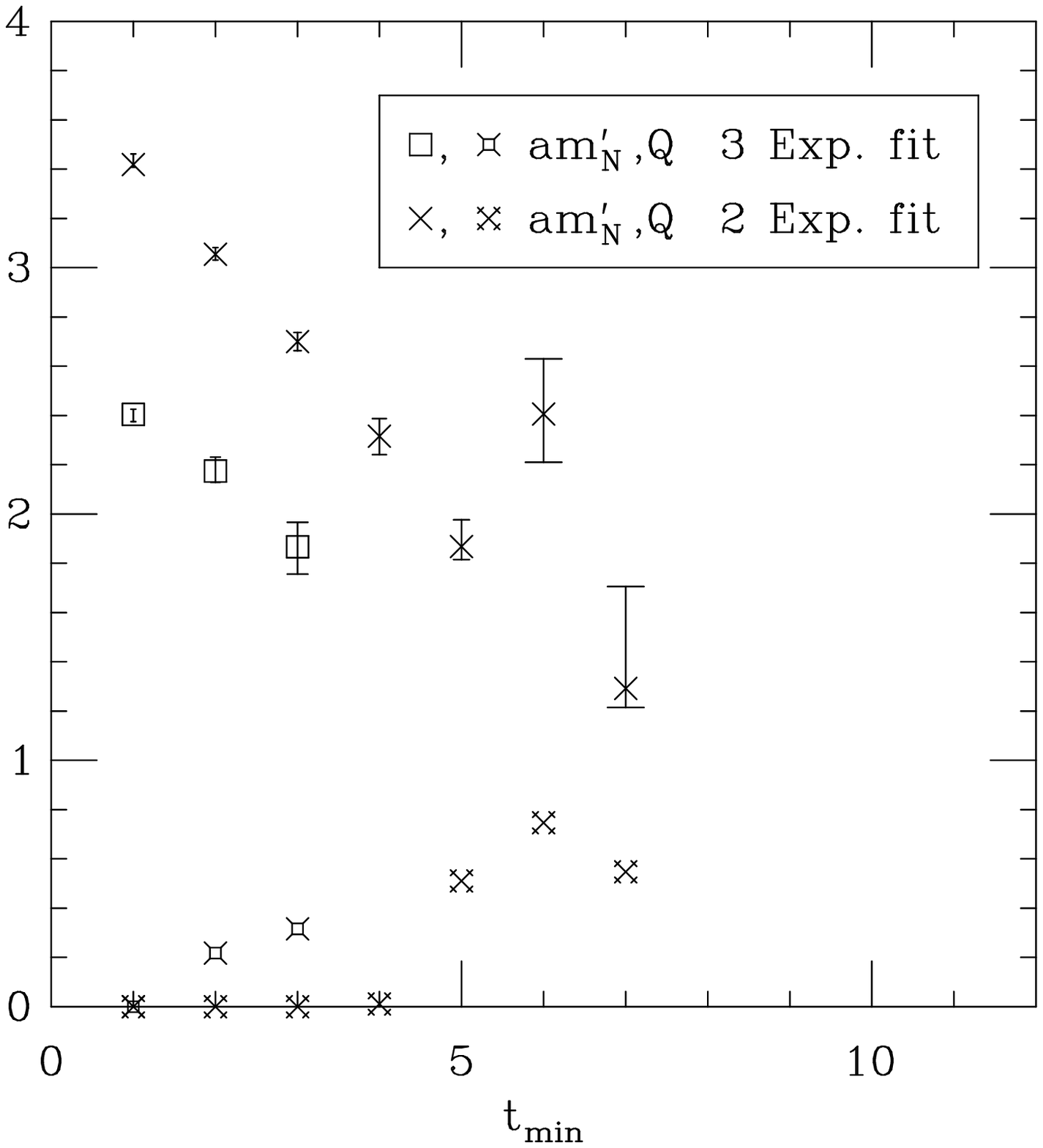}
\caption{$am_{N}$, $am'_{N}$ and $Q$   versus
$t_{min}$ for $c=1.57$,  $12^3 \times 24$, $\kappa = 0.14077$,
using  local and smeared propagators.}
\label{fig:nuc-57-multi-fit}
\end{figure}

As can be seen in Fig.~(\ref{fig:nuc-57-multi-fit}) there is
significant second and even third state contamination for the nucleon
when local and smeared operators are used in the fit.
Hence only those correlators calculated with smeared operators,
with overlap onto the $J^P=1/2^-$ state, were used to determine
$am_N$.
The contribution of the parity partner of Eq.(\ref{eq:baryon-twopt}) was found
to be sufficiently suppressed if $t_{max}$ was chosen to be $T - 1$.
The fit ranges, using a 2--exponential fit, were 2--11 and 2--15.

In the case of the $\Delta$,
the higher state contamination was not as large
as for the nucleon. Therefore
local and  smeared operators were used. The fit ranges
were 5--11 and 5--15 with a 2--exponential fit.

\section{Results}

The masses obtained for the pseudoscalar, vector, nucleon and $\Delta$
for each value of the clover coefficient and combination of quark masses,
are listed in
Table~\ref{mps-c-coarse} to Table~\ref{mdelta-c-coarse}.
The larger lattice size
corresponds to one
used by the GF11 collaboration
with the Wilson fermion action and the same
$\beta$~\cite{GF11}, so that we are also able
to compare results for non-zero $c$ with those for $c=0$.
One expects  the effect of changing $c$ will be
more noticeable at our coarse lattice spacing
than at a larger $\beta$.
The effect of reducing the physical
volume to $12^3 \times 24$ was also investigated, using the
tadpole-improved SW action.

\subsection{Effect of clover coefficient}

\subsubsection{The chiral limit}
For small quark masses, the bare mass of a quark on the lattice
can be defined as
\begin{equation}
a m_q = \frac{1}{2} \left( \frac{1}{\kappa} - \frac{1}{\kappa_{\mathrm{crit}}}
\right) \; ,
\label{eq:pole-mass-def}
\end{equation}
where $\kappa_{\mathrm{crit}}$ is {\it a priori} an undetermined
function of $\beta$.
We use the standard extrapolation  in quark mass for
pseudoscalar mesons, neglecting possible logarithmic
divergences described by Sharpe \cite{sharpe-logs},
\begin{equation}
(am_{PS})^2
 = b_\kappa + \frac{c_\kappa}{\kappa} + O(\kappa^{-2}) \;\; ,
\label{eq:m-ps-vs-kappa}
\end{equation}
where
\begin{equation}
\kappa_{\mathrm{crit}} = -\frac{c_\kappa}{ b_\kappa} \;\; .
\end{equation}
However, as noted by Bhattacharya {\it et al.} \cite{LANL} and Collins
{\it et al.} \cite{SCRI}, the terms which are $O(\kappa^{-2})$
 cannot be entirely neglected for
the quark masses used in this study.
A linear extrapolation in $1/\kappa$ leads to a large $\chi^2/\rm{d.o.f.}$, as
can be seen in Table~\ref{kappa-crit}.  An estimate of the systematic
uncertainty was obtained by performing a quadratic fit through the
three masses and a linear fit to the two lightest masses.  In all the
cases considered, the deviation from the original linear fit was
greater for the quadratic fit than for the linear fit to the two
lightest masses.  The systematic error quoted in 
Table~\ref{kappa-crit} is
conservatively estimated to be the deviation
of the quadratic fit from the original linear fit.

We note that the value for $\kappa_{\mathrm{crit}}$ is always larger when the
quadratic form is employed, regardless of the clover coefficient
or lattice size used.
Hence, results for other
observables will always be quoted with an entirely positive or
negative systematic error.

As can be seen from  Table~\ref{kappa-crit} 
(including the GF11\cite{GF11}
data for comparison),
$\kappa_{\mathrm{crit}}$ approaches  $1/8$ as $c$ is increased from $0$ to $1$
and that $\tilde{\kappa}_{\mathrm{crit}}$ in the tadpole improved case is  closer still.

\subsubsection{Meson masses}
In this section, the physical pseudoscalar and vector masses are evaluated
by extrapolation and interpolation in the quark masses to the
appropriate physical values.
Certain input parameters are necessary to do this.
In particular, for mesons containing up and down
valence quarks
(which are assumed to be degenerate in mass and
will be referred to here as ``normal''),
one may use the experimental values for
$M_\pi$ and $M_\rho$
(we apply a convention that experimentally
determined masses are labelled with an  ``M'', while
those calculated on the lattice are labelled with an
``m'').
Effectively, one of these  sets the
quark mass while the other sets the lattice spacing.

The vector mass extrapolation has the following form
\begin{equation}
am_V = am_\rho^{\mathrm{crit}} + c_V (am_{PS})^2 + O((am_{PS})^3) \;\; ,
\label{eq:mv-vs-mpssq}
\end{equation}
where logarithmic terms due to the quenched approximation have
been discarded.
The constant term $am_\rho^{\mathrm{crit}}$ corresponds to the vector
mass in the chiral limit.
Following the procedure outlined by the GF11 collaboration,
values of $am_\pi$ and $am_\rho$ are determined using the
physical ratio
\begin{equation}
\label{eq:pi-rho-phys}
\frac{am_\pi}{am_\rho} = \frac{M_\pi}{M_\rho} = 0.1792  \;\; .
\end{equation}
%

\begin{figure}
\pspicture{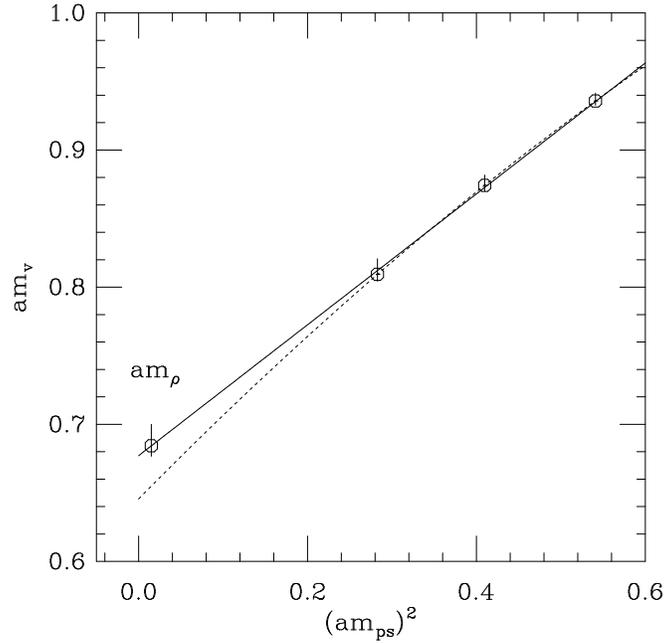}
\caption{$am_V$ as a function of $(am_{PS})^2$,
 $N_s^2 \times N_t = 16^3 \times 32$, $c=1.57$.
The solid line indicates the central value for
the  linear fit to all three masses. The  dashed line is
the quadratic fit. The calculated value for $am_\rho$
for the linear fit is
also quoted.}
\label{fig:mv-vs-mpisq}
\end{figure}

Once again, the systematic error due to higher order corrections
is estimated by quadratically fitting all three masses
and performing a linear fit in the two lightest masses.
The deviation due to the quadratic fit was again found
to be consistently larger.
An example of this is shown in Fig.~(\ref{fig:mv-vs-mpisq}).
The resulting values for $am_\rho$ (including the GF11 \cite{GF11}
data) are quoted in
Table~\ref{chiral-mesons}.
Having used the ratio of Eq.(\ref{eq:pi-rho-phys}) to fix the normal
quark mass, the scale can be determined using either $m_\pi$ or $m_\rho$.

\begin{figure}
\pspicture{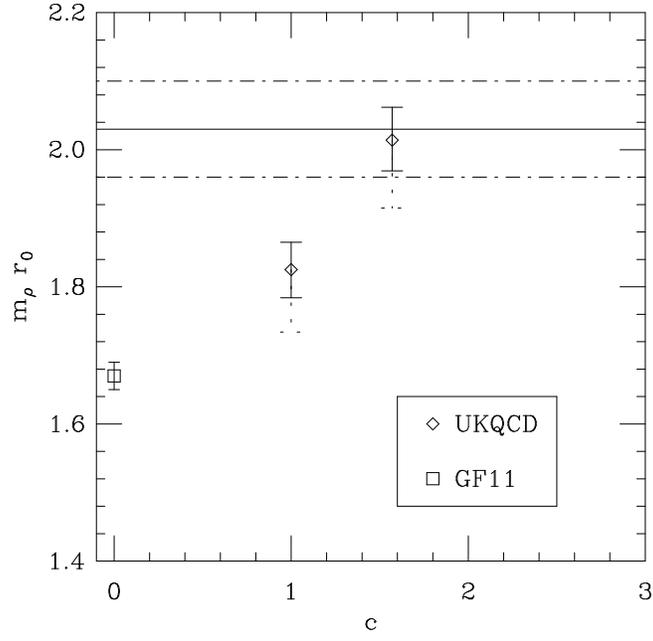}
\caption{$m_\rho r_0$ versus $c$.
Statistical errors on the data points
are marked with solid lines.
Systematic errors due to the quadratic chiral
extrapolation are marked, on the data points,  with dashed  lines.
The horizontal lines indicate the  continuum
limit from the GF11 data (a finite--volume correction
has been included), along with the statistical error
of the fit to the continuum.
Systematic effects due to discretisation errors in
$r_0$, which are $O(a^2)$
(necessary for the extrapolation to the continuum),
have {\it not} been included.
}
\label{fig:mrhor0-vs-c}
\end{figure}

It is useful to compare $m_\rho$ with the lattice measurement of a
gluonic quantity, where discretisation errors are $O(a^2)$
 and hence can
be expected to be smaller.  We choose Sommer's
force parameter, $r_0$ \cite{sommer}.  We
can extrapolate the GF11 values for
$m_\rho r_0$ versus $ar_0^{-1}$ to the continuum limit
which yields
\begin{equation}
\left.{m_\rho} {r_0}\right\vert^{Quenched}_{a=0} = 2.03 \pm 0.07 \;\; .
\label{eq:extrap-ratio}
\end{equation}
This includes a  correction which the GF11 collaboration have used
to eliminate finite volume effects, which rounds
the result down by approximately 4\%.
Assuming that $r_0$ and the string tension, $\sqrt{K}$ are related
by $r_0 \sqrt{K} = 1.18$
and interpolating the available string tension data 
from $\beta=5.7-6.5$, one finds
$r_0/a$ at $\beta=5.7$ to be  $2.94$.
One can then compare our
data for $m_\rho r_0$ at $\beta=5.7$ as a function of $c$
with the continuum limit from GF11.
These results are  plotted  as a function of $c$
in Fig.~(\ref{fig:mrhor0-vs-c}),
noting that there are significant
discretisation effects in the force parameter at $\beta=5.7$
which have not been  taken into account.
There is a clear trend toward the continuum limit
as the clover coefficient is increased to its tadpole improved
value.

The determination of
meson masses containing strange valence
quarks requires as input the
experimental mass of a strange meson,
for example $M_K$.
With this mass as input, one can determine
$am_K$ by requiring :
\begin{equation}
\frac{am_K}{am_\rho} = \frac{M_K}{M_\rho} = 0.643 \;\; .
\label{eq:amK-from-ratio}
\end{equation}
{}From the condition of
Eq.(\ref{eq:amK-from-ratio}) and employing 
Eq.(\ref{eq:mv-vs-mpssq}),
one can then predict $am_{K^*}$ fixed from $am_K$,
which we refer to as  $am_{K^*}(am_K)$.
Our results for $am_{K^*}(am_K)$ and the ratio
$(am_{K^*}(am_K))/{am_\rho}$
can be found in columns 3 and 4 of 
Table~\ref{strange-masses}.
We note that the ratio $(am_{K^*}(am_K))/{am_\rho}$
at $c=1$ is consistent to within 1 standard
deviation with that
at $c=1.57$ and that the central value
lies several standard deviations below the
experimental value.
There are large  systematic errors
due to the chiral extrapolation at both
values of the clover coefficient, however
this error is smaller
than the difference between our
results and experimental data.
The discrepancy in this  ratio
has also been noted at $\beta=6.0$,
with $c=0$ by Bhattacharya {\it et al.} \cite{LANL}.

The choice of strange meson
is not unique. Instead, one could have
fixed $am_{K^*}$ from
\begin{equation}
\frac{am_{K^*}}{am_\rho} = \frac{M_{K^*}}{M_\rho} = 1.160\;\; ,
\label{eq:amKstar-from-ratio}
\end{equation}
and through Eq.(\ref{eq:mv-vs-mpssq}) one can then predict
$am_K$ fixed from $am_{K^*}$, which we refer to as
$am_K(am_{K^*})$.
Our results at both clover coefficients for this mass and the ratio
$(am_K(am_{K^*}))/{am_\rho}$
are listed in columns
5 and 6 of Table~\ref{strange-masses}.
We note  that the ratio $(am_K(am_{K^*}))/{am_\rho}$
is also constant to within one standard deviation as $c$ is changed
from 1 to $1.57$ and that the central value
lies several standard deviations above the
experimental value.
However, in this case, the systematic errors
due to the chiral extrapolation at both
values of the clover coefficient are so large
that we cannot demonstrate that
these ratios  are inconsistent with experiment.

The mass $am_\phi$
of the pure valence ${\overline{s}s}$ vector state can be determined
similarly, but a valence ${\overline{s}s}$ pseudoscalar,
$\eta_s$, is not observed.
However, using an estimate of
$M_{\eta_s}$ by
Lipps {\it et al.} \cite{lipps}, we can estimate
the ratio of these masses :
\begin{equation}
\frac{M_\phi}{\mbox{``}M_{\eta_s}\mbox{''}}
\approx \frac{am_V(\overline{s}s)}{am_{PS}(\overline{s}s)}
= 1.5 \;\; .
\label{eq:mvss-mpss-ratio}
\end{equation}

It is therefore possible  to determine $am_V(\overline{s}s)$,
from Eq.(\ref{eq:mv-vs-mpssq}) and Eq.(\ref{eq:mvss-mpss-ratio})
without extrapolating to the chiral limit, which
we have seen previously, has large systematic errors.
The resulting masses are shown in Table~\ref{phi-masses}.

\begin{figure}
\pspicture{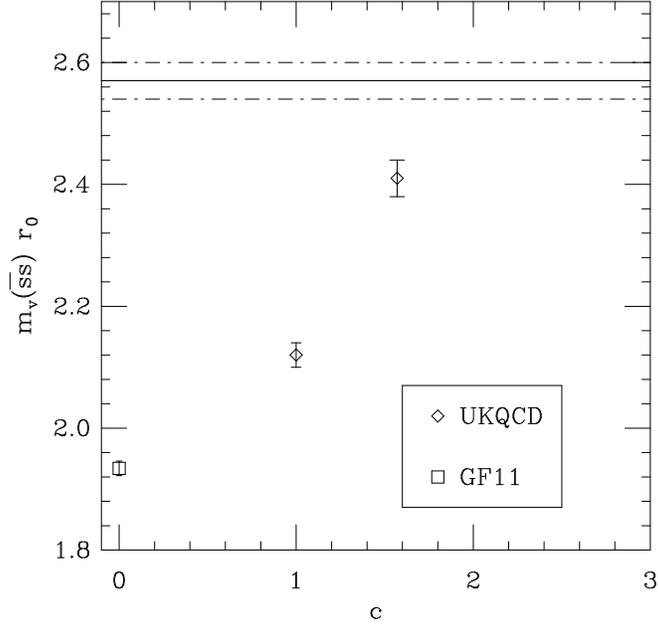}
\caption{$m_V(\overline{s} s) r_0$ versus $c$.
The horizontal line indicates the
continuum limit  from the GF11 data
 ( a finite--volume correction
 has been included ).
Systematic effects due to discretisation errors in
$r_0$, which are $O(a^2)$
(necessary for the extrapolation to the continuum),
have {\it not} been included.
}
\label{fig:mphir0-vs-c}
\end{figure}

Using the data from the GF11 collaboration,
it is possible to calculate $am_V(\overline{s}s)$
for $c=0$ for $\beta=5.7$ and the other gauge couplings.
Assuming a linear behaviour with respect to
the lattice spacing, the
continuum limit of $m_V(\overline{s}s) r_0$
using the GF11 data has  been evaluated.
It should be noted, however that the linear
extrapolation in the lattice spacing for the
GF11 data is very poor, having a $\chi^2/\rm{d.o.f.}$ of
approximately 13, even though the fit is
uncorrelated.
It is likely therefore  that the
continuum limit for $m_V(\overline{s} s) r_0$
has a large systematic error due to this fit.
There is also a correction to
infinite volume which shifts the
value downwards.
The behavour of $m_V(\overline{s} s) r_0$
with respect to $c$ at $\beta=5.7$
is shown in Fig.~(\ref{fig:mphir0-vs-c}).
The absence of the systematic error
due to the chiral extrapolation demonstrates the effect
of the clover coefficient  more clearly than from
$m_\rho r_0$.
Again, we find there is a clear trend toward the continuum limit
as the clover coefficient is increased to its tadpole improved
value.

\subsubsection{Mass splittings}
Heavy Quark Effective Theory (HQET)
predicts that for heavy--light mesons,
the vector-pseudoscalar mass splitting,
$\Delta_{\mathrm{V-PS}} = m_V^2 - m_{PS}^2$, is constant.
This is borne out by experiment, with
$M_{D*}^2 - M_{D}^2 \approx 0.53 \; \mbox{GeV}^2$
and $M_{B*}^2 - M_{B}^2 \approx 0.49 \; \mbox{GeV}^2$.
A somewhat unexpected experimental result is that
this trend is continued into the
light quark regime,
where the
hyperfine splitting,
$\Delta_{\mathrm{V-PS}}$,
remains  approximately
constant at $0.55 \; \mbox{GeV}^2$.

Quenched lattice simulations fail noticeably
to  reproduce this behaviour.
HQET predicts that  $\Delta_{\mathrm{V-PS}}$ is proportional
to $\langle \overline{h} \sigma_{\mu\nu}F^{\mu\nu} h \rangle$,
where $h$ is the heavy quark field.
As the clover term is of this form,
naively one would then  expect that increasing the
size of clover coefficient  would
reduce this discrepancy at least for heavy--light
systems.
Tentative comparisons with the $c=0$ and $c=1$ actions at
$\beta=6.2$ with low statistics indicated that the
fall off in the splitting had decreased \cite{ukqcd:npblight}.

\begin{figure}
\pspicture{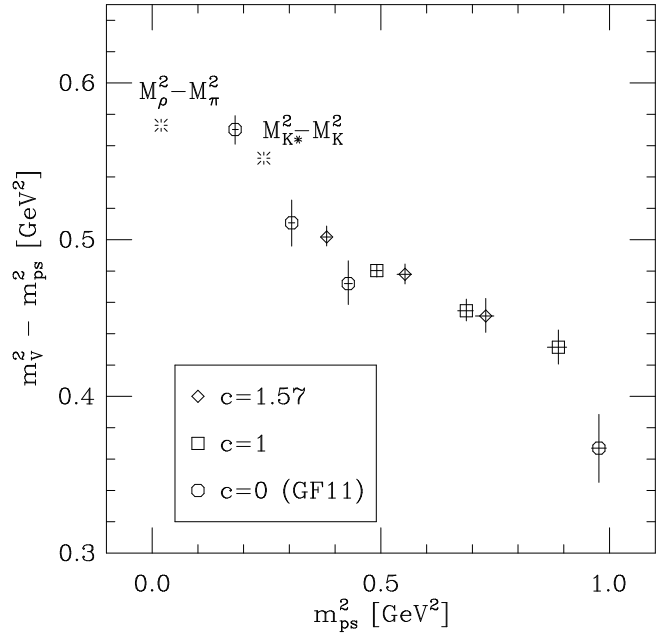}
\caption{The hyperfine splitting $m_V^2 - m_{PS}^2$ versus
$m_{PS}^2$ for all the three values of $c$ at $\beta=5.7$, $16^3 \times 32$.}
\label{fig:mvsq-mpssq-comp}
\end{figure}

In Fig.~(\ref{fig:mvsq-mpssq-comp}) the splittings
from the three different values of the clover coefficient are compared.
The scale for each action is chosen from
$M_{K^*}$.
The slope $\partial{(a^2 \Delta_{\mathrm{V-PS}})}/\partial{(am_{PS})^2}$
is unaffected by this choice.
While there is a noticeable change in the slope
on going from $c=0$ to $c=1$, the slopes at
$c=1$ and $c=1.57$ are consistent with each other.
The remaining discrepancy is presumably due to the
error of the quenched approximation.

\subsubsection{The $J$ Parameter}
\begin{figure}
\pspicture{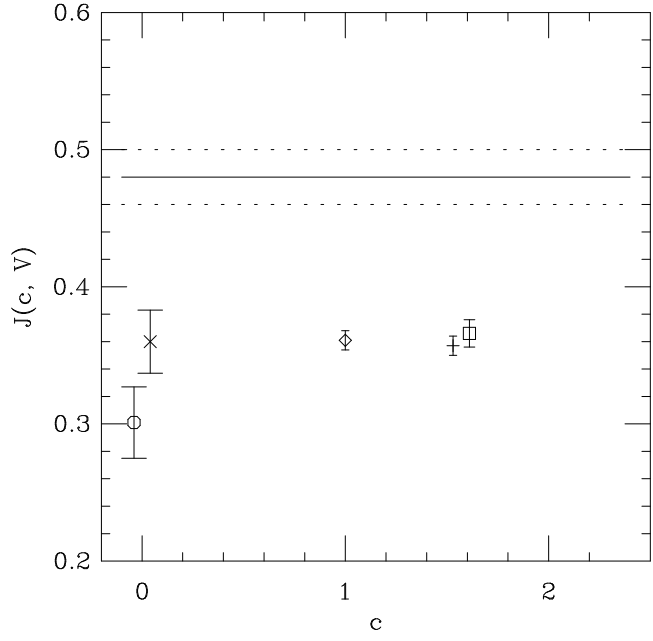}
\caption{The parameter $J$ against $c$ for all
values of $c$ and volumes.
The horizontal line indicates the ``experimental value'';
GF11, $c=0$,  $16^3 \times 32$ ($\circ$);
GF11, $c=0$,  $24^3 \times 32$ ($\times$);
UKQCD, $c=1.0$,  $16^3 \times 32$, ($\diamond$);
UKQCD, $c=1.57$, $16^3 \times 32$, ($\Box$);
UKQCD, $c=1.57$, $12^3 \times 24$
($+$).
}
\label{fig:Jplot}
\end{figure}

As noted previously,
it is useful to be able to compare lattice spectrum results with
existing experimental data without an extrapolation to the chiral limit.
The parameter $J$, defined as\cite{UKQCD:J}
\begin{equation}
J \equiv m_{K^*}\frac{dm_V}{dm_{PS}^2} \;\; , \;\; \frac{m_{K^*}}{m_K} = 1.8
\;\; .
\end{equation}
allows such a comparison.
Existing quenched Wilson--like fermion actions  yield  values around
$J = 0.37$ whereas
an estimate of $J$ using experimental data yields
$ J = 0.48(2)$.
In Fig.~(\ref{fig:Jplot}),  
$J$ versus $c$ (including the calculated
value of J at two volumes from the GF11 collaboration)
is plotted.
We find
\begin{eqnarray}
J(\beta=5.7, c=1, 16^3 \times 32) &=& 0.361 \pm 7 \;\; , \nonumber \\
J(\beta=5.7, c=1.57, 16^3 \times 32) &=& 0.366 \pm 10 \;\; .
\label{eq:J-bigvol}
\end{eqnarray}

The values of $J$ from Eq.(\ref{eq:J-bigvol}) and 
Eq.(\ref{eq:J-smallvol})
below,
for both non--zero values
 of $c$ and both volumes, agree with the world average of the
quenched data, and disagree with the experimental estimate.
It should be noted that $J$ is trivially related to the slope
$\partial{(a^2 \Delta_{\mathrm{V-PS}})}/\partial{(am_{PS})^2}$
outlined in the previous section.
We therefore expect that the prescription
that solves the anomalous behaviour
of $\Delta_{\mathrm{V-PS}}$ will also solve the disagreement in
$J$.

\subsubsection{Baryons}
We extrapolate the nucleon mass to the normal-quark
limit assuming a linear dependence on the quark mass :
\begin{equation}
a m_N = a m_N^{crit} + c_N (am_{PS})^2 + O((am_{PS})^2) \;\;.
\end{equation}
We extrapolate the  $\Delta$ mass likewise.
The final results for the nucleon and $\Delta$
are quoted in Table~\ref{chiral-nucleon} and 
Table~\ref{chiral-delta} respectively.

\begin{figure}
\pspicture{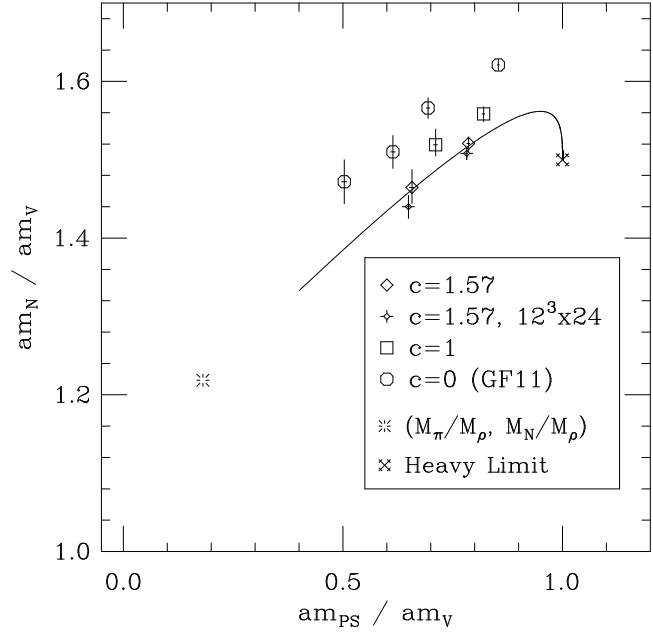}
\caption{The Edinburgh plot for all three values of $c$.}
\label{fig:edin-plot-fiveseven}
\end{figure}

{}From the combined results for the pseudoscalar, vector and
nucleon masses, we show the  ``Edinburgh'' plot in
Fig.~(\ref{fig:edin-plot-fiveseven}).
One finds a statistically significant
difference between the ratios at each value of $c$.
As $c$ is increased, the trend of the data is towards
the phenomenological curve of Ono \cite{ono-phenom}.
Furthermore, the ratio $m_N/m_\rho$
approaches the experimental value $M_N/M_\rho$,
but even at $c=1.57$ is still approximately
$13\%$ too large.

\subsection{Finite volume effects}
\begin{figure}
\pspicture{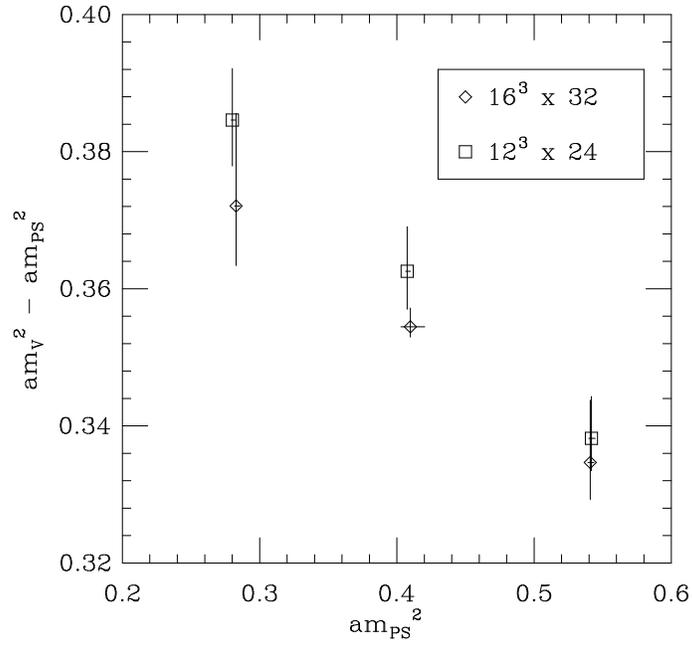}
\caption{$(am_V)^2 - (am_{PS})^2$ plotted against $(am_{PS})^2$
for $c=1.57$,
 $N_s^2 \times N_t = 16^3 \times 32$ and $ 12^3 \times 24$.}
\label{fig:hyperfine-volume}
\end{figure}

The masses obtained for the $12^3 \times 24$ lattice
are listed  in Table~\ref{mps-c-coarse} to 
Table~\ref{mdelta-c-coarse}.
As stated previously, it proved to be somewhat more difficult
to extract reliable masses for this volume.
As before, $\tilde{\kappa}_{\mathrm{crit}}$ is evaluated with a statistical and systematic
error to be
\begin{equation}
\tilde{\kappa}_{\mathrm{crit}}(\beta=5.7, c=1.57, 12^3 \times 24) = 
0.12348^{+ 2}_{- 2} + 24
\;\; ,
\end{equation}
which agrees with the result from the larger volume and has
a similarly sized systematic error.
Likewise, as shown in 
Fig.~(\ref{fig:hyperfine-volume}),
the hyperfine splittings are consistent  to
within 1 standard deviation.
The chirally
extrapolated and strange  meson masses
are determined as in section III.A.2  and
the results listed in Table~\ref{chiral-mesons} and 
Table~\ref{strange-masses}.
Once again, the results are consistent to
within one standard deviation with those on the larger volume.
Similarly, the parameter $J$ is determined to be
\begin{equation}
J(\beta=5.7, c=1.57, 12^3 \times 24) = 0.357 \pm 7 \;\; ,
\label{eq:J-smallvol}
\end{equation}
which is consistent with the larger volume.

Both baryons are more strongly affected by the size of the lattice.
The nucleon masses  at both $\kappa$'s are approximately
two standard deviations smaller than in the larger volume.
This is somewhat unexpected as other studies in quenched
QCD using Wilson--like fermions indicate that the
nucleon mass falls with increasing size over a similar
range of volumes (using $am_\rho$ to determine the
lattice spacing, we see that our volumes vary
from $(2 \, \mbox{fm})^3$ to $(2.6 \, \mbox{fm})^3$
approximately).
We note that the $Q$ values for the fit to the nucleon
masses on the $12^3 \times 24$ lattice are very close to 1, which
may indicate that the statistical errors are underestimated.
We find the extrapolated value
\begin{equation}
a m_N(\beta=5.7,c=1.57,12^3 \times 24) = 0.931^{+ 15}_{- 15}  \;\; ,
\end{equation}
which is also two standard deviations smaller than in the
larger volume.
The $\Delta$ masses at both $\kappa$'s lie
approximately two standard deviations
above the values on the larger lattice,
and
\begin{equation}
a m_\Delta(\beta=5.7,c=1.57,12^3 \times 24)
= 1.167^{+ 22}_{- 23}  \;\; .
\end{equation}

\section{CONCLUSIONS}
In this paper, we have examined the effect, at $\beta = 5.7$,
of changing the clover coefficient and volume on the quenched
light-hadron spectrum computed using the SW fermion action.
As the  clover coefficient is increased
there is better agreement
between the perturbative (tree--level)
 and non-perturbative calculation of $\kappa_{\mathrm{crit}}$.
When the clover coefficient is changed
from $c=0$ (the Wilson action) to  $c=1$ (the SW action)
and $c=1.57$ (the tadpole--improved SW action)
there is a significant difference in the masses of the
chirally extrapolated and strange pseudoscalar and vector mesons,
in the nucleon and $\Delta$ masses
 and  in  the Edinburgh plot.

Interestingly, a number of other  quantities,
for example $m_V^2  -  m_{PS}^2$, $J$ and
the ratios $am_K/am_\rho$ and  $am_{K^*}/am_\rho$
do not appear to change significantly
as $c$ is changed from $1.0$ to $1.57$.
As the finite volume effects appear to be under control,
and these observables have been chosen to avoid
the systematic errors due to the chiral extrapolation,
the possible remaining systematic errors
are the effect of quenching
the gauge configurations and a possibly large deviation of the mean field
estimate of the clover coefficient from $c_{\rm non-pert}$.
It would therefore be very interesting then to examine the
behaviour of these quantities in any future
studies in full QCD
under changes in the value of the clover coefficient.

In changing the volume from approximately $(2 \, \mbox{fm})^3$ to
$(2.6 \, \mbox{fm})^3$
the mesonic observables are consistent to within
one standard deviation.
Baryon masses are consistent to within two standard deviations.
Unfortunately, with this data, one cannot differentiate
between  different
{\it Ans\"atze} used for describing the
volume behaviour of  masses
\cite{big-luescher}, \cite{fukugita}.


The Alpha
collaboration \cite{alpha}, \cite{alpha-louis}
has calculated the
clover coefficient non--perturbatively
for $6.0 \le \beta \le 6.8$.
In general, the coefficients obtained through
 this approach are significantly larger than
those obtained via tadpole--improvement,
although the coefficients converge
as $\beta$ is increased.
Our data appears to suggest
that $c_{\rm non-pert}$ could at $\beta=5.7$
be somewhat larger than the tadpole improved value.

Currently, we are carrying out an analysis of the quenched light hadron
mass spectrum
at $\beta=6.0$ and $\beta=6.2$
using the tadpole improved SW action\cite{rdk-louis}.
This will directly explore whether better scaling is achieved
using $c=1/u_0^3$ than with $c=1$.

\section{Acknowledgements}
HPS would like to thank the staff at
Brookhaven National Laboratory for
their kind hospitality during his stay there.
DGR acknowledges the
support of PPARC through an Advanced Fellowship, and the support of
Argonne National Laboratory during the completion of this work.
The UKQCD collaboration wish to express their thanks to the Edinburgh
Parallel Computing Centre for their support and
maintenance of the Meiko Computing Surfaces.

%
%

\begin{center}
\begin{table}
\begin{tabular}{ c c c c c }
 & State & $J^P$ & Correlators & $\Gamma$ structure \\ \hline
Mesons & PS & $0^-$ & $\left\langle{P(t) P^\dagger(0)}\right\rangle$
 & $P = \overline{\psi}^a \gamma_5 \psi^a$\\
\hline
 & V & $1^-$ & 
$\left\langle{\vec{V}_1(t) \cdot \vec{V}_1^{\dagger}(0)}\right\rangle$,
& $\vec{V}_1 = \overline{\psi^a} \vec{\gamma}\psi^a$\\
 &  &  & $\left\langle{\vec{V}_2(t) \cdot \vec{V}_2^{\dagger}(0)}\right\rangle$
& $\vec{V}_2 = \overline{\psi^a} \vec{\gamma} \gamma_4 \psi^a$ \\\hline
Baryons & N & $\frac{1}{2}^-$ & 
$\left\langle{N_1(t) \overline{N}_1(0)}\right\rangle$,
& $N_1 = \varepsilon_{abc} (\psi^a C \gamma_5 \psi^b)\psi^c$\\
 &  &  & 
$\left\langle{N_2(t)\overline{N}_2(0)}\right\rangle$
 & $N_2 = \varepsilon_{abc}(\psi^a C \gamma_4 \gamma_5 \psi^b)\psi^c$ \\ \hline
 & $\Delta$ & $\frac{3}{2}^-$ &
$\left\langle{\Delta(t) \overline{\Delta}(0)}\right\rangle$,
& $\Delta = \varepsilon_{abc}(\psi^a C \gamma_\mu \psi^b) \psi^c$\\
\end{tabular}
\caption{Hadron operators. The quark fields may be smeared and
there is an implicit sum over spatial sites. Lower case latin variables
indicate colour indices.}
\label{operator-list}
\end{table}

\end{center}

\begin{center}
\begin{table}
\begin{tabular}{ c c c c c c c c c }
$c$ &  $N_s^3 \times N_t$  &
 $\kappa_1$ & $\kappa_2$ & $am_{PS}$ &  Fit Range  & $\chi^2$ &$\rm{d.o.f.}$& $Q$  \\
\hline
1.57 & $12^3 \times 24$ &
 0.13843 & 0.13843 & $0.7361^{+ 18}_{- 14}$ &  3--12 & 14.8 & 22 & 0.872  \\
\hline
1.57 & $12^3 \times 24$ &
 0.14077 & 0.13843 & $0.6384^{+ 18}_{- 8}$ &  3--12 & 15.6 & 22 & 0.834  \\
\hline
1.57 & $12^3 \times 24$ &
0.14077 & 0.14077 & $0.5292^{+ 21}_{- 8}$ &  3--12 & 17.7 & 22 & 0.724 \\
\hline
\hline
1.0 &  $16^3 \times 32$ &
0.14663 & 0.14663 & $0.7343^{+ 19}_{- 13}$ &  3--16 & 33.9 & 34  & 0.474  \\
\hline
1.0 & $16^3 \times 32$ &
0.14948 & 0.14663 & $0.6458^{+ 22}_{- 12}$ &  3--16 & 37.1 & 34 & 0.326  \\
\hline
1.0 & $16^3 \times 32$ &
 0.14948 & 0.14948 & $0.5462^{+ 25}_{- 12}$ &  3--16 & 43.8 & 34 & 0.121 \\
\hline
\hline
1.57 & $16^3 \times 32$ &
 0.13843 & 0.13843 & $0.7355^{+ 18}_{- 12}$ &  3--16 & 43.0 & 34 & 0.139  \\
\hline
1.57 & $16^3 \times 32$ &
0.14077 & 0.13843 & $0.6402^{+ 22}_{- 12}$ &  3--16 & 44.8 & 34 & 0.102  \\
\hline
1.57 & $16^3 \times 32$ &  0.14077 & 0.14077 & 
$0.5319^{+ 27}_{- 11}$ &  3--16 &
53.6 & 34 & 0.017 \\
\end{tabular}
\caption{Pseudoscalar masses at both volumes and values of $c$.}
\label{mps-c-coarse}
\end{table}
\end{center}

\begin{center}
\begin{table}
\begin{tabular}{ c c c c c c c c c }
$c$ & $N_s^3 \times N_t$ &
 $\kappa_1$ & $\kappa_2$ & $am_{V}$ &  Fit Range  & $\chi^2$ &$\rm{d.o.f.}$& $Q$  \\
\hline
1.57 & $12^3 \times 24$ &
 0.13843 & 0.13843 & $0.9381^{+ 33}_{- 20}$ &  4--12 & 46.2  & 40 & 0.230 \\
\hline
1.57 & $12^3 \times 24$ &
 0.14077 & 0.13843 & $0.8775^{+ 39}_{- 26}$ &  4--12 & 39.6 & 40 & 0.488  \\
\hline
1.57 & $12^3 \times 24$ &
 0.14077 & 0.14077 & $0.8153^{+ 50}_{- 38}$ &  4--12 & 39.6 & 40 & 0.488  \\
\hline
\hline
1.0 & $16^3 \times 32$ & 0.14663 & 0.14663 & $0.8950^{+ 41}_{- 21}$
 &  4--16 & 24.2  & 31
& 0.802  \\
\hline
1.0 & $16^3 \times 32$ & 0.14948 & 0.14663 & $0.8325^{+ 51}_{- 23}$
 &  4--16 & 20.9 & 31 &
0.913 \\
\hline
1.0 & $16^3 \times 32$ & 0.14948 & 0.14948 & $0.7680^{+ 61}_{- 35}$
 &  4--16 & 22.0 & 31 &
0.882  \\
\hline
\hline
1.57 & $16^3 \times 32$ &
 0.13843 & 0.13843 & $0.9357^{+ 52}_{- 28}$ &  4--16 & 21.7  & 31 & 0.893 \\
\hline
1.57 & $16^3 \times 32$ &
 0.14077 & 0.13843 & $0.8743^{+ 64}_{- 38}$
 &  4--16 & 22.7 & 31 & 0.861  \\
\hline
1.57 & $16^3 \times 32$ &
 0.14077 & 0.14077 & $0.8093^{+ 91}_{- 50}$
 &  4--16 & 24.0 & 31 & 0.811  \\
\end{tabular}
\caption{Vector masses at both volumes and values of $c$.}
\label{mv-c-coarse}
\end{table}
\end{center}

\begin{center}
\begin{table}
\begin{tabular}{ c c c c c c c c }
$c$ & $N_s^3 \times N_t$ &
 $\kappa_1$ & $am_{N}$ &  Fit Range  & $\chi^2$ &$\rm{d.o.f.}$& $Q$  \\
\hline
1.57 &  $12^3 \times 24$ &
 0.13843 & $1.4147^{+ 62}_{- 53}$
 &  2--11 & 2.0  & 14 & 0.999 \\
\hline
1.57 &  $12^3 \times 24$ &
 0.14077 & $1.1741^{+ 106}_{- 94}$ &  2--11 & 6.4 & 14 & 0.955  \\
\hline
\hline
1.0 & $16^3 \times 32$ & 0.14663 & $1.3948^{+ 96}_{- 71}$ 
&  2--15 & 20.3  & 22  & 0.564
\\
\hline
1.0 &  $16^3 \times 32$ & 0.14948 & $1.1667^{+ 159}_{- 84}$
 &  2--15 & 25.4 & 22 & 0.279
\\
\hline
\hline
1.57 &  $16^3 \times 32$ & 0.13843 & $1.4231^{+ 87}_{- 79}$
 &  2--15 & 17.6  & 22 &
0.728 \\
\hline
1.57 &  $16^3 \times 32$ &
0.14077 & $1.1853^{+ 187}_{- 116}$
 &  2--15 & 23.1 &  22 & 0.397  \\
\end{tabular}
\caption{Nucleon masses at both volumes and values of $c$.}
\label{mn-c-coarse}
\end{table}
\end{center}

\begin{center}
\begin{table}
\begin{tabular}{ c c c c c c c c }
$c$ & $N_s^3 \times N_t$ &
$\kappa_1$ & $am_{\Delta}$ &  Fit Range  & $\chi^2$ &$\rm{d.o.f.}$& $Q$  \\
\hline
1.57 & $12^3 \times 24$ &
 0.13843 & $1.5447^{+ 87}_{- 68}$ &  5--11 & 8.3  & 8 & 0.409 \\
\hline
1.57 & $12^3 \times 24$ &
0.14077 & $1.3564^{+ 156}_{- 127}$ &  5--11 & 3.2 & 8 & 0.922  \\
\hline
\hline
1.0 & $16^3 \times 32$ &
0.14663 & $1.4834^{+ 112}_{- 57}$ &  5--15 & 11.0  & 16  & 0.808  \\
\hline
1.0 & $16^3 \times 32$ &
0.14948 & $1.2812^{+ 180}_{- 92}$ &  5--15 & 11.4 & 16 & 0.782  \\
\hline
\hline
1.57 & $16^3 \times 32$ &
 0.13843 & $1.5251^{+ 125}_{- 80}$ &  5--15 & 10.4  & 16 & 0.845 \\
\hline
1.57 & $16^3 \times 32$ &
0.14077 & $1.3167^{+ 207}_{- 143}$ &  5--15 & 16.1 &  16 & 0.445  \\
\end{tabular}
\caption{$\Delta$ masses at both volumes and values of $c$.}
\label{mdelta-c-coarse}
\end{table}
\end{center}
\begin{center}
\begin{table}
\begin{tabular}{ c l c c }
$N_s^3 \times N_t$ & $c$ & $\kappa_{\mathrm{crit}}$ 
& $\chi^2/\rm{d.o.f.}$   \\
\hline
$12^3 \times 24$ & 1.57
& $0.123480^{+ 15}_{- 15} + 238$ & 23.0  \\
\hline
\hline
$16^3 \times 32$ & 0.0 (GF11) & $0.169405\; \pm \;52$ (Stat). &  - \\ \hline
$16^3 \times 32$ & 1.0 & $0.153184^{+ 37}_{- 38} + 268$ &  4.5  \\ \hline
$16^3 \times 32$ & 1.57 & $0.123466^{+ 27}_{- 26} + 176$ &  5.5
\end{tabular}
\caption{Results for $\kappa_{\mathrm{crit}}$ 
($\tilde{\kappa}_{\mathrm{crit}}$ in the tadpole improved
case), including the GF11 data at this $\beta$ for comparison.
In the case of the UKQCD data,
the first error quoted is statistical  and the second is the systematic
shift due to the fit to a quadratic form. The value of $\chi^2/\rm{d.o.f.}$
quoted is for the linear fit.}
\label{kappa-crit}
\end{table}
\end{center}
\begin{center}
\begin{table}
\begin{tabular}{ c l c c c c  }
$N_s^3 \times N_t$ & $c$ & $a m_\pi$ & $a m_\rho^{crit}$ &
$a m_\rho$  &  $\chi^2/\rm{d.o.f.}$  \\
\hline
$12^3 \times 24$ & 1.57 & $0.1250^{+ 14}_{- 13} - 59$ &
$0.6897^{+ 77}_{- 75} - 349$ &
$0.6969^{+ 77}_{- 74} - 327$ & 6.7  \\
\hline
\hline
$16^3 \times 32$ & 0.0 (GF11)  & -  &
- &
$0.5676 \pm 79$ (Stat.) & - \\ \hline
$16^3 \times 32$ & 1.0 & $0.1113^{+ 15}_{- 15} - 56$  &
$0.6143^{+ 85}_{- 86} - 325$ &
$0.6208^{+ 84}_{- 86} - 309$ & 2.4  \\ \hline
$16^3 \times 32$ & 1.57 & $0.1228^{+ 20}_{- 19} - 60$  &
$0.6778^{+ 113}_{- 106} - 357$  &
$0.6850^{+ 113}_{- 106} - 336$ &  2.1  \\
\end{tabular}
\caption{Chirally extrapolated results for the vector and pseudoscalar,
at the zero and normal quark mass limits.
The GF11 data at this $\beta$
is included for comparison.
Another
value for $am_\rho$ was also computed by GF11 using a different
smearing radius which is approximately 1--2 standard deviations
smaller than the one quoted here.
In the case of the UKQCD data, the first error quoted is statistical  and
the second is the systematic
shift due to using quadratic chiral extrapolations.
The value of $\chi^2/\rm{d.o.f.}$ quoted is for the linear fit.}
\label{chiral-mesons}
\end{table}
\end{center}

\begin{center}
\begin{table}
\begin{tabular}{ c c c c  c c }
$N_s^3 \times N_t$ & $c$ &
$am_{K^*}(am_K)$ &
$(am_{K^*}(am_K))/{am_\rho}$ &
$am_{K}(am_{K^*})$ &
$(am_K(am_{K^*}))/{am_\rho}$
  \\
\hline
$12^3 \times 24$ & 1.57
& $0.782 \pm 7 - 18$ & $1.122 \pm 2 + 28$ &
$0.505 \pm 9 - 65$ & $0.724 \pm 6 - 94$
 \\
\hline
\hline
$16^3 \times 32$ & 1.0 & $0.698 \pm 9 - 21$ &
$1.124\pm 2 + 24$
& $0.447 ^{+ 8}_{- 9} - 53$ & $0.721 \pm 6 - 85$
\\ \hline

$16^3 \times 32$ & 1.57 &
$0.771 ^{+ 11}_{- 10} - 21$ & $1.125 \pm 3 + 28$ &
$0.491^{+ 13}_{- 12} - 60$ &
$0.717 \pm 8 - 88$ \\
\end{tabular}
\caption{Results for the strange mesons
 at both volumes and $c$, using $M_K$
(columns 3 and 4) and $M_{K^*}$
(columns 5 and 6) to fix the strange quark mass.
The first error quoted is statistical  and
the second is the systematic
shift due to the use of a quadratic chiral extrapolation. }
\label{strange-masses}
\end{table}
\end{center}

\begin{center}
\begin{table}
\begin{tabular}{ c c c c c }
$N_s^3 \times N_t$ & $c$ &
$am_V(\overline{s} s )$ &
$am_{PS}(\overline{s} s )$ &
$\chi^2/\rm{d.o.f.}$  \\
\hline
$12^3 \times 24$ & 1.57
& $0.831 \pm 6$ & $0.554 \pm 4$ &
 6.7 \\
\hline
\hline
$16^3 \times 32$ & 1.0 & $0.742^{+ 7}_{- 8}$ &
$0.495 \pm 5$
& 2.4 \\ \hline

$16^3 \times 32$ & 1.57 &
$0.821^{+ 10}_{- 9}$ &
$0.547^{+ 7}_{- 6}$  &
 2.1 \\
\end{tabular}
\caption{Masses for  the valence $\overline{s}s$ states, defined from
fixing the ratio $am_V(\overline{s} s )/am_{PS}(\overline{s} s) $ to  $1.5$.
}
\label{phi-masses}
\end{table}
\end{center}

\begin{center}
\begin{table}
\begin{tabular}{ c l l l }
$N_s^3 \times N_t$ & $c$ & $a m_N^{crit}$ &
$a m_N(a m_\pi)$ \\
\hline
$12^3 \times 24$ & 1.57 & $0.916^{+ 16}_{- 15}$ &
$0.931^{+ 15}_{- 15} $   \\
\hline
\hline
$16^3 \times 32$ & 1.0 & $0.891^{+ 23}_{- 24}$ & 
$0.902^{+ 23}_{- 23}$
\\ \hline
$16^3 \times 32$ & 1.57 & $0.931^{+ 30}_{- 28}$
& $0.945^{+ 29}_{- 28}$  \\
\end{tabular}
\caption{Chirally extrapolated results for the nucleon,
at the zero and normal quark mass limit.
The $\chi^2/\rm{d.o.f.}$
is not quoted
as the number
of degrees of freedom is equal to the number of data points. }
\label{chiral-nucleon}
\end{table}
\end{center}

\begin{center}
\begin{table}
\begin{tabular}{ c l l l }
$N_s^3 \times N_t$ & $c$ & $a m_\Delta^{crit}$ &
$a m_\Delta(a m_\pi)$ \\
\hline
$12^3 \times 24$ & 1.57 &$1.156^{+ 23}_{- 24}$ &
$1.167^{+ 22}_{- 23}$  \\
\hline
\hline
$16^3 \times 32$ & 1.0 & $1.036^{+ 25}_{- 24}$ & 
$1.047^{+ 25}_{- 24}$
\\ \hline
$16^3 \times 32$ & 1.57 & $1.091^{+ 30}_{- 30}$
& $1.103^{+ 29}_{- 30}$  \\
\end{tabular}
\caption{Chirally extrapolated $\Delta$ masses,
at the zero and  normal quark mass limit.
The $\chi^2/\rm{d.o.f.}$
is not quoted
as the number
of degrees of freedom is equal to the number of data points. }
\label{chiral-delta}
\end{table}
\end{center}

%
%

%
%

\begin{figure}
\pstwopictures{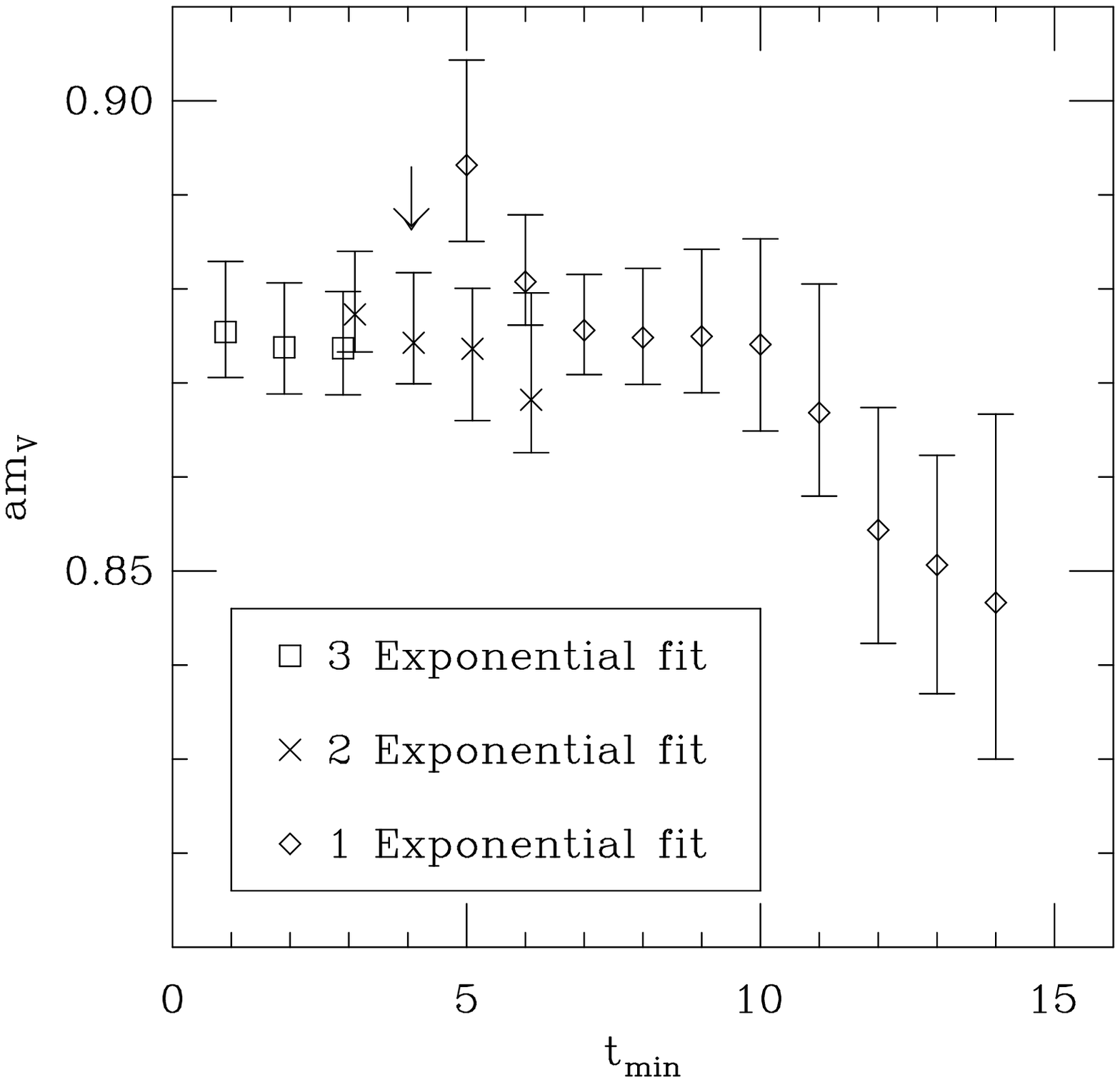}{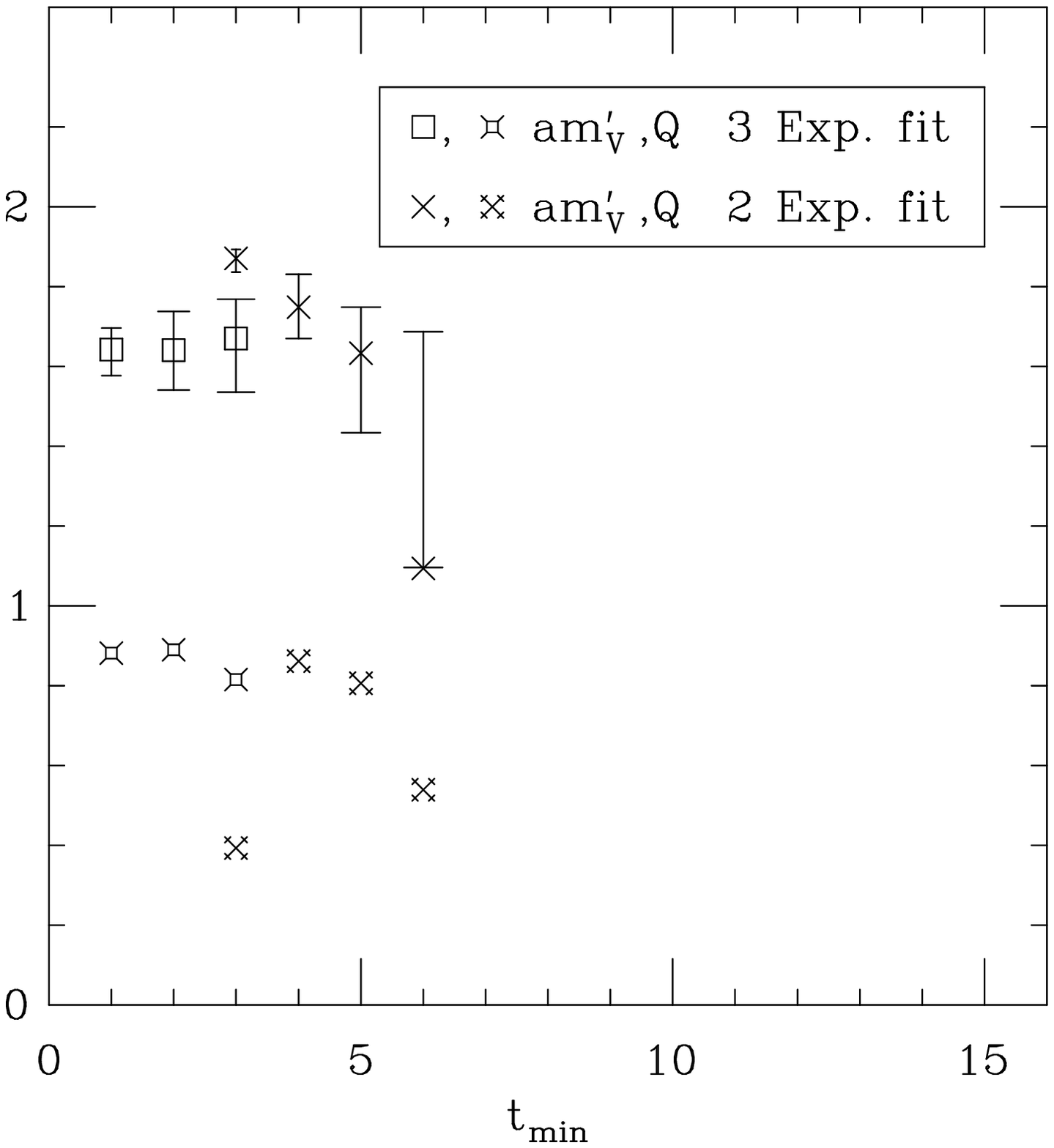}
\caption{$am_{V}$, $am'_{V}$ and $Q$    versus
$t_{min}$ for $c=1.57$ at $16^3 \times 32$, $\kappa _1= 0.14077$,
$\kappa _2= 0.13843$
using local and smeared propagators.
 The arrow indicates which fit range was used for the final result.}
\label{fig:v-57-multi-fit}
\end{figure}

\begin{figure}
\pstwopictures{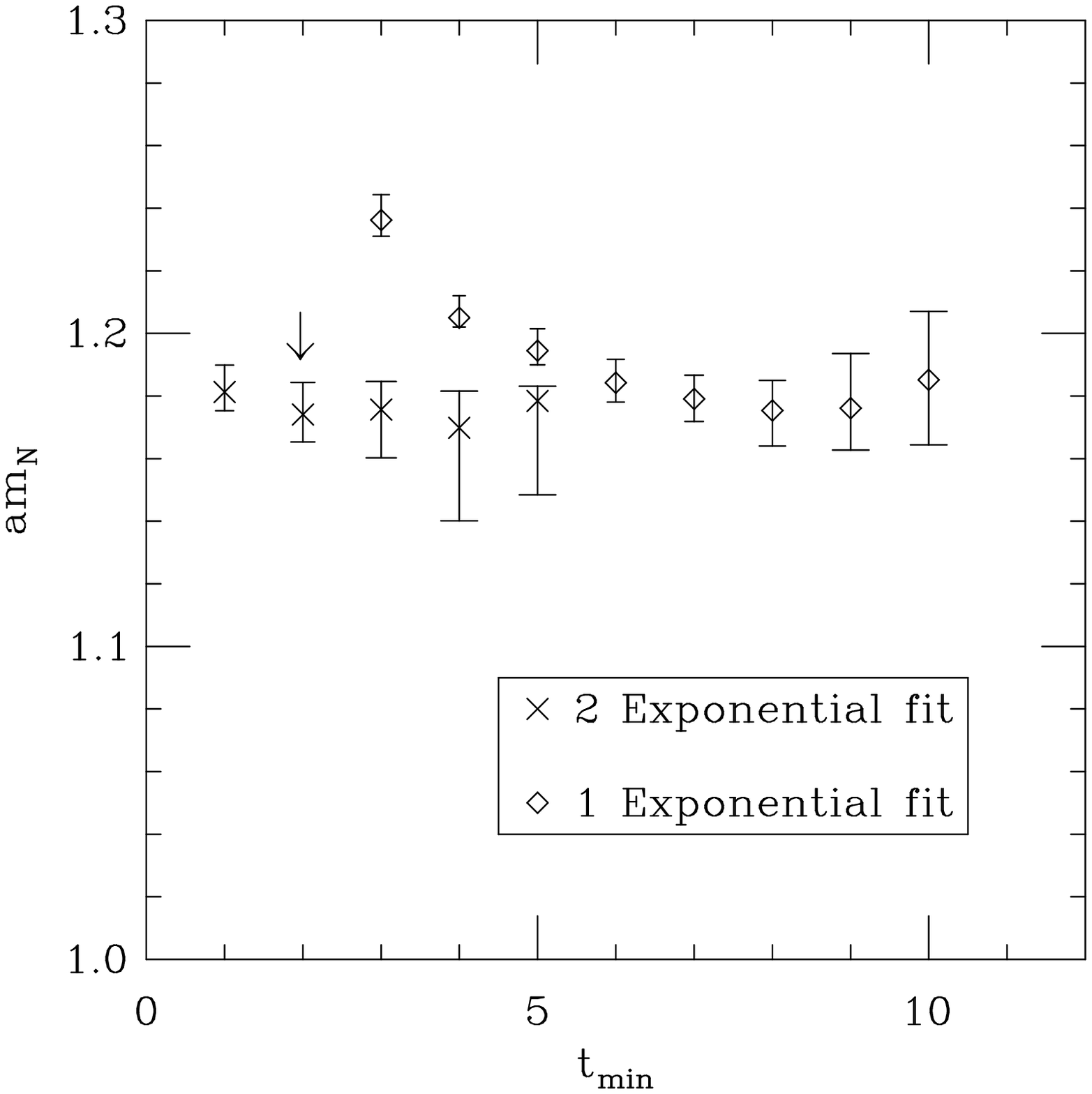}{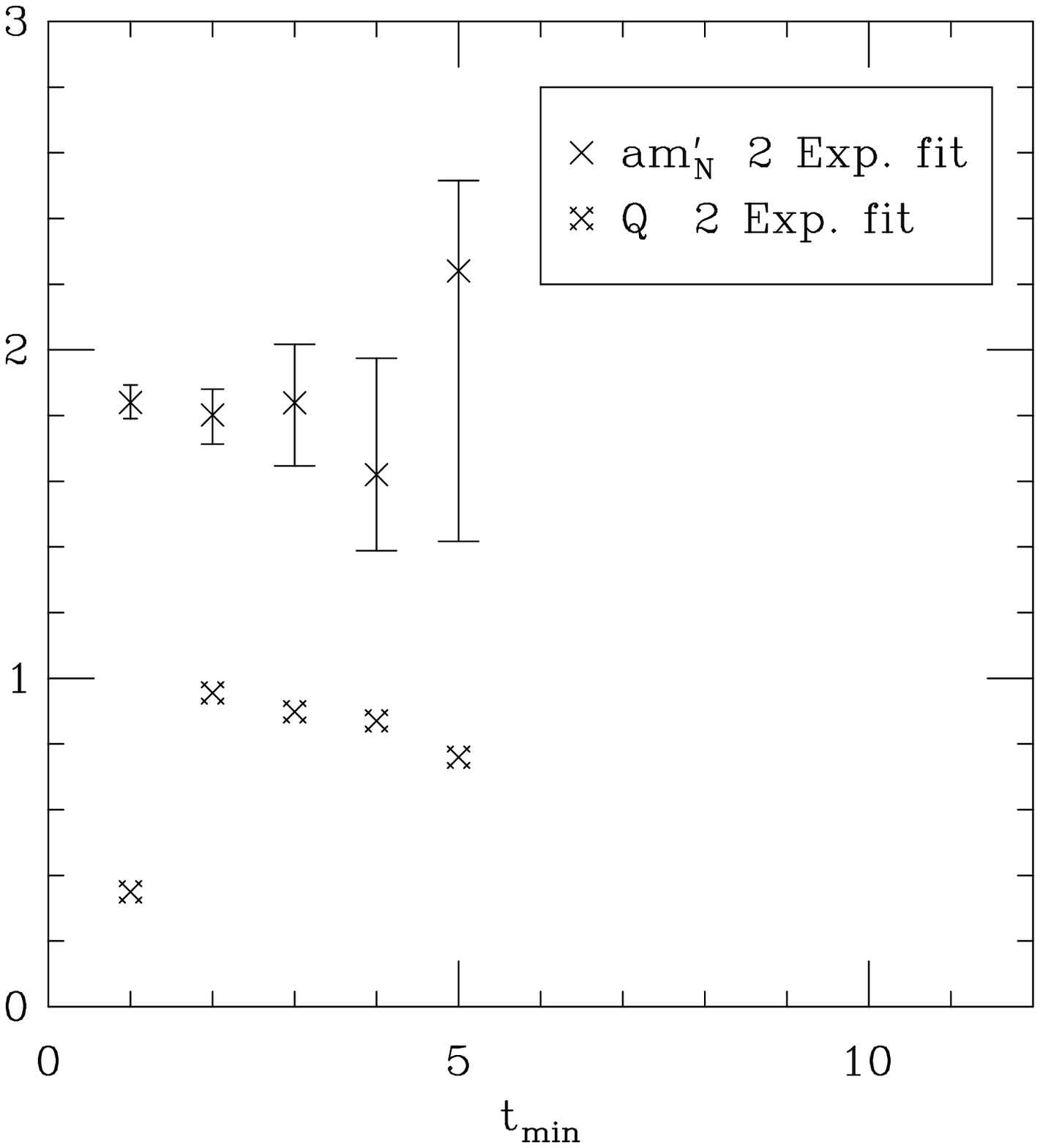}
\caption{$am_{N}$, $am'_{N}$ and $Q$   versus
$t_{min}$ for $c=1.57$, $12^3 \times 24$, $\kappa = 0.14077$,
using only  smeared propagators.
 The arrow indicates which fit range was used for the final result.}
\label{fig:proton-57-SSS}
\end{figure}

\end{document}